\documentclass{epl}
\usepackage{amssymb}
\title{Stick-slip instability for viscous fingering in a gel}
\shorttitle{Stick-slip instability...}
\author{N. Puff\inst{1} \and D. Bonn\inst{2} \and \\ G. Debr{\'e}geas\inst{1} \and J.-M. di Meglio\inst{1}\thanks{
Universit{\'e} Louis Pasteur and Institut Universitaire de France} \and  D. Higgins\inst{1} \and C. Wagner\inst{2}}
\shortauthor{N. Puff \etal}
\institute{
  \inst{1} Institut Charles Sadron (CNRS UPR22), 6 rue Boussingault,
67083 Strasbourg Cedex, France\\
  \inst{2} Laboratoire de Physique Statistique (CNRS UMR 8550), \'Ecole Normale Sup{\'e}rieure,
24 rue Lhomond, 75231 Paris Cedex 05, France
}

\author{N. Puff \and  G. Debr{\'e}geas \and J.-M. di Meglio\thanks{
Universit{\'e} Louis Pasteur and Institut Universitaire de France} \and  D. Higgins (\inst{1}), \\
D. Bonn \and C. Wagner (\inst{2})}
\shortauthor{N. Puff \etal}
\institute{
  \inst{1} Institut Charles Sadron (CNRS UPR22), 6 rue Boussingault,
67083 Strasbourg Cedex, France\\
  \inst{2} Laboratoire de Physique Statistique (CNRS UMR 8550), \'Ecole Normale Sup{\'e}rieure,
24 rue Lhomond, 75231 Paris Cedex 05, France
}

      
\begin{document}

\maketitle

\begin{abstract}
The growth dynamics of an air finger injected in a visco-elastic gel
(a PVA/borax aqueous solution) is studied in a linear Hele-Shaw cell.
Besides the standard Saffmann-Taylor instability, we observe - with increasing
finger velocities - the existence of two new regimes: (a) a stick-slip
regime for which the finger tip velocity oscillates between 2 different values,
producing local pinching of the finger at regular intervals, (b) a
``tadpole'' regime where a fracture-type propagation is
observed. A scaling argument is proposed to interpret the dependence
of the stick-slip frequency with the measured rheological properties
of the gel. 
\end{abstract}

\section{Introduction}
When a fluid is injected into a more viscous liquid, the invading front is unstable at early stages 
and splits into multiple
fingers. This process, known as the Saffmann-Taylor (ST) problem\cite{st},
has been extensively studied as a simple example of non-linear dynamical
instability. The width of the fingers arises from
a competition between the surface tension, which tends to widen them,
and the viscous stress, which tends to make them thinner. In a standard ST experiment, the fluids are confined between two parallel plates separated by a gap of thickness $b$ much smaller than any other length scale. For air
 driven into a Newtonian liquid, the stationary  width \( w \) of the fingers relative to the gap thickness $b$ only depends
on the capillary number \( Ca=\eta U/\gamma  \) where \( \eta  \)
is the liquid viscosity, \( U \) the bubble tip velocity and \( \gamma  \)
the liquid surface tension. When the bubble is forced to grow in a
linear channel of width $W<<b$, the ratio \( w/W \) decreases as \( Ca \) is increased until it reaches
a limiting value equal to one-half.

In most practical situations however, the liquid phase (typically
polymer solutions or complex multi-phase systems) does not show a
purely Newtonian rheology. Many experimental and theoretical studies
have thus attempted to understand the modifications of the ST results
associated with more complicated rheological characteristics of the
invaded phase\cite{maher,coussot,lindner1}. Two situations may be distinguished. For liquids with
shear rate dependent viscosity, the shape of the finger is slightly
modified. This can be understood by considering the inhomogeneous
shearing experienced by the liquid across the channel : the shear
rate in the central part (of order \( \dot{\gamma }\sim U/b \)
with \( b \) being the plate spacing of the Hele-Shaw cell) is higher
than on the sides. Hence, a bubble penetrating a shear-thinning liquid
for instance tends to be thinner than what is expected
for a Newtonian liquid because the central part of the channel offers
less resistance to the flow\cite{lindner2,shelley}. For strongly elastic 
liquids, a richer variety
of situations is observed because elastic stresses can develop in
the liquid phase and affect the bubble growth. This internal
stress generally enhances tip splitting or even leads to fracture
formation, yielding highly ramified patterns\cite{maher,zhao, vlad}.

Here we study the rapid penetration of a gas into a visco-elastic gel
which exhibits a plateau in the stress/strain rate curve. Three different
tip dynamics are evidenced. A low velocity regime where the bubble
shape remains very similar to the standard ST process. A high velocity
regime where the bubble assumes a tadpole profile. In between
this regime, we observe a stick-slip transition regime where the
bubble tip velocity oscillates between two different values. Stick
to slip transitions are found to occur at regular spatial intervals
and produce local pinchings of the finger.

\section{Experiments}

We have used aqueous solutions of PVA/Borax as model visco-elastic gels. PVA (poly(vinyl alcohol)) has been purchased from Prolabo (Rhodoviol 30/5) and borax (sodium tetraborate decahydrate) from Aldrich. We have followed the preparation method described by Garlick\cite{garlick}. The visco-elasticity of this
 solution arises from temporal cross-linkings of the
PVA chains through H-bonding with the borax groups. The concentration of PVA in water was fixed and equal to 3 \% (weight/weight) while the concentration of borax has been changed between 0.125 and 5 \% (w/w) in order to vary the rheological properties.

The Hele-Shaw is made of two rectangular plastic (PMMA) plates of thickness 
1.25 cm, tightened together by 12 screws. The two plates are separated by a thin
Teflon sheet of uniform thickness \( b \). This spacer is U-shaped
to define a one-end open channel of width \( W \) and length $L$ = 20 cm
filled with the gel. In most experiments, the distance \( b \) and
\( W \)  are respectively \( 1 \) and 5 mm (in a few experiments, \( W \)
and \( b \) have been changed to check for geometrical constraints effects).
 A 1 mm hole is drilled in the lower plate to allow
for the air injection. The air is supplied by a 25 ml syringe
at constant pressure \( P \) measured by a gauge in derivation.
A bright background is placed below the cell to give a good
contrast on the bubble edge. Images are captured in the central part
of the set-up to avoid end effects with a high speed video camera
(Kodak Motion Corder). Subsequent tracking of the bubble tip is performed
using an image analysis software (IDL) to produce growth velocity
sequences \( U(t) \). In the standard ST regime, the tip velocity
\( U \) is proportional to the applied pressure gradient \( (P-P_{0})/x \)
where \( x \) is the distance to the outlet and \( P_{0} \) the
external pressure. In a single experiment, we therefore explore
a large range of velocities \( U \). We furthermore vary \( (P-P_{0}) \)
(around $1$ atmosphere) to access a wider dynamical range.

\section{Results}

\begin{figure}
\onefigure[scale=0.4]{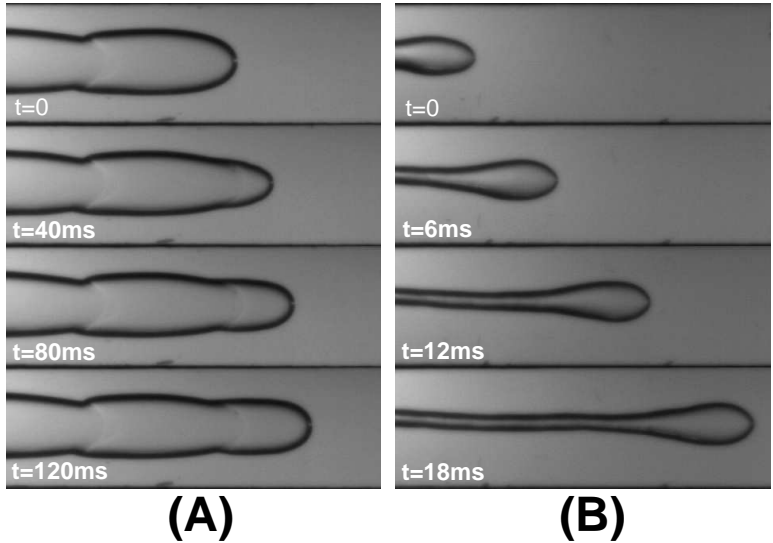}
\caption{Sequences of growth showing the two new regimes. The width \(W\) 
and depth \(b\) of the channel are respectively 5 and 1 mm.
 (A): Stick-slip regime. After bulging, the deposited 
film is thicker: the shape of the bubble in the first image can 
therefore be detected on the following snapshots. (B): The rapid ``tadpole''
regime.}
\label{f.1}
\end{figure}

At low velocity, the bubble grows smoothly with a velocity \( U \)
increasing as the inverse of the distance to the outlet. The width
\( w \) of the bubble however is found to be slightly below the value
\( W/2 \) expected for a Newtonian liquid. For a certain threshold
velocity (of the order of a few mm.s\( ^{-1} \)), the bubble experiences
a sudden bulging (see Fig. \ref{f.1}A). The tip hops rapidly ahead,
with a significant decrease of its radius of curvature. In about \( 30 \)
ms, the velocity increases by a factor of \( 10 \) to \( 20 \). The
velocity then rapidly recovers its initial value as the bubble front
relaxes to its initial shape, leaving a pinch, a narrow section of the finger behind (Fig. \ref{f.1}A). We also observe,
with the bright background illumination, that the shape of the bubble
at the onset of bulging remains visible after the tip has moved ahead.
This indicates that the deposited liquid film thickness is strongly
modified at this position: the pinching also occurs in the vertical
direction.

\begin{figure}
\onefigure[scale=0.5]{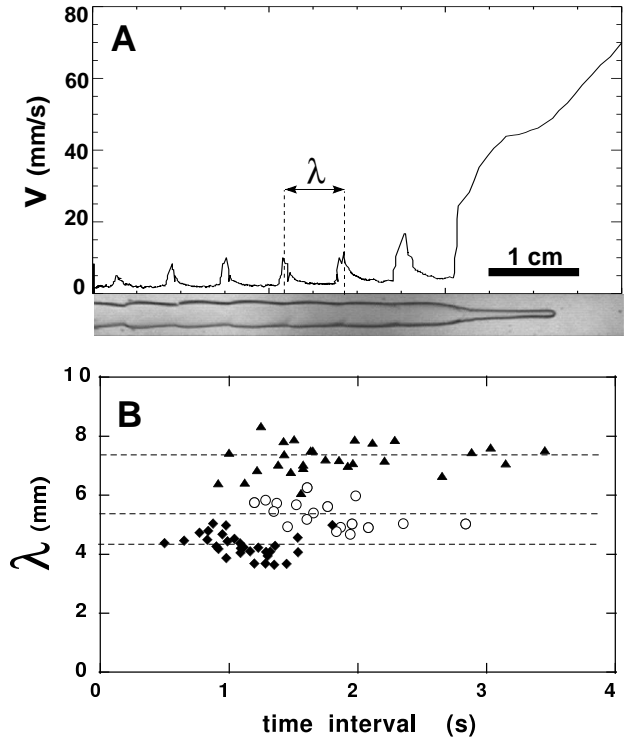}
\caption{Periodicity of the bulgings in the stick slip regime. (A) Tip 
velocity as a function of its position in the channel (the snapshot 
corresponds to the bubble shape after the end of the stick-slip
regime). Each bulging
is associated with an overshoot of the tip velocity. The maxima are used
 to evaluate the stick-slip wavelength. The last bulge
 is followed by the ``tadpole'' regime where the tip velocity rapidly
increases. (B) Different measurements of the stick-slip wavelength as a
function of the time interval between the two tip velocity overshoots.
Each set of data corresponds to a different gel (Borax concentration
are respectively  $\blacktriangle$: 0.125 \%, 
{\large $\circ$}: 1 \%, $\blacklozenge$: 2 \%). For each gel,
the stick-slip wavelength is independent of the time interval between
bulgings. }
\label{f.2}
\end{figure}

Once the first bulging is triggered, we observe up to fifteen other events
occurring at regular distances \( \lambda  \) (Figure \ref{f.2}). Figure \ref{f.2}B
shows the different values of \( \lambda  \) for different gels as
a function of the time interval \( \Delta t \) between the 2 successive bulges.
Each experiment provides several measurements of \( \lambda  \) over
a certain range of \( \Delta t \). This range is then extended
by varying the applied pressure, yielding the large number of data
points. For a given gel, \( \lambda  \) appears to be independent
of \( \Delta t \), or equivalently of the tip velocity at the onset of bulging.
However, \( \lambda  \) decreases as the borax volume fraction is
increased.

In all experiments, the last bulge is followed by a transition to
a regime of rapid bubble growth (see Fig. \ref{f.1}A): 
instead of recovering its original
velocity and shape, the tip keeps accelerating and the bubble assumes
a tadpole shape, characterized by a rounded head and a thin cylindrical
tail (of diameter smaller than the gap $b$).

\section{Rheology}\label{rheo}

To characterize the visco-elastic properties of the gels, several
standard rheological measurements have been performed. The rheometer
in use was a Reologica Stresstech with cone and plate geometry (diameter
40 mm, angle 4$^\circ$). These measurements were checked against
those in a Couette cell: no significant differences between
the two geometries were observed, validating the results.

\begin{figure}
\twofigures[scale=.6]{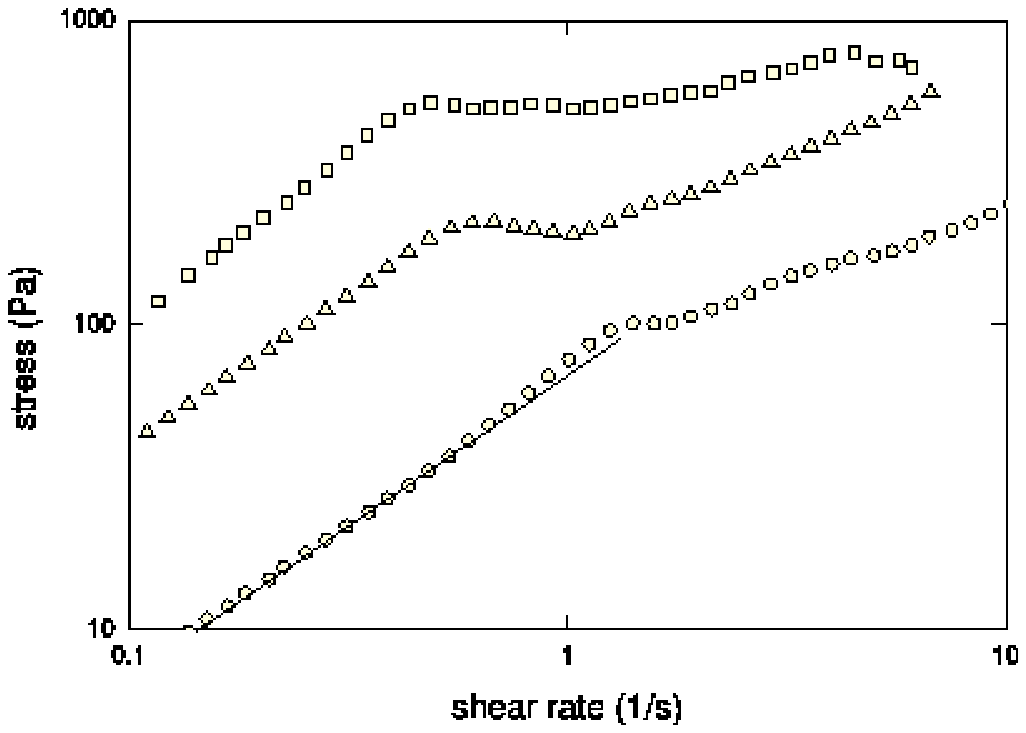}{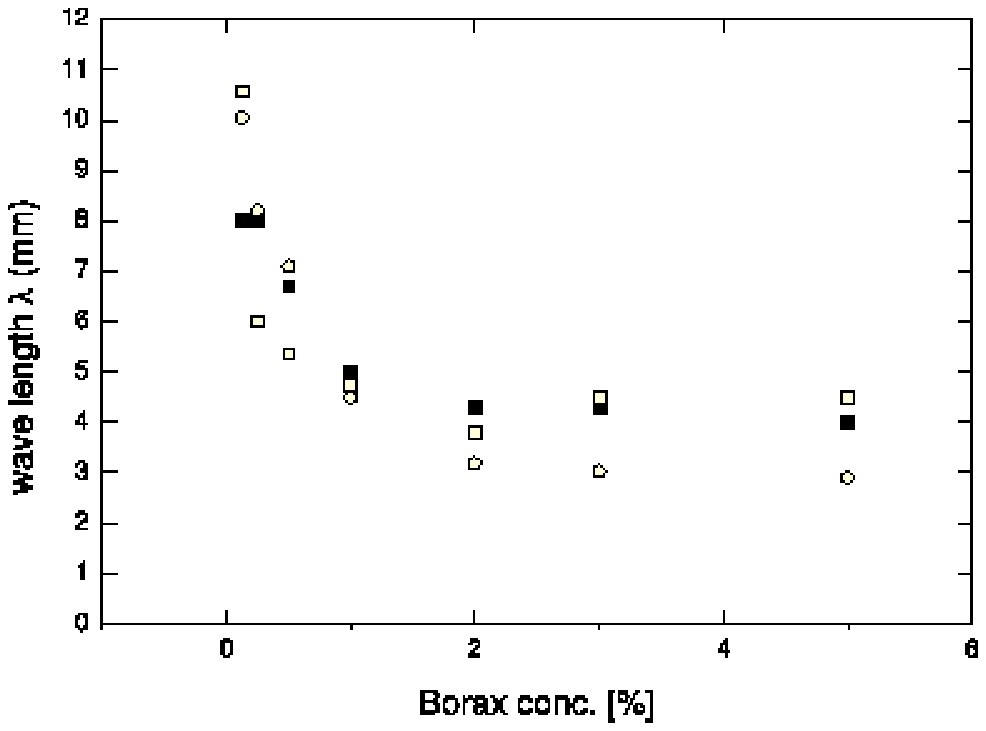}
\caption{Shear stress \(\sigma\) as a function of the shear rate 
$\gamma$ for different borax concentrations. {\large $\circ$}: 0.25~\%, 
$\triangle$: 0.5 \%, $\square$: 1 \%. The straight line is 
the extrapolated linear fit of the 0.25 \% borax data at shear rates
$\gamma < 0.5$ s$^{-1}$.}
\label{f.3a}
\caption{Stick-slip wavelength $\lambda$ from: ($\blacksquare$)
 the ST experiment, calculated with equation
 \ref{equation} using the plateau stress $\sigma_p$ ({\large $\circ$}) 
or the elastic modulus $G_0$ deduced from the oscillatory measurements 
($\square$). The numerical prefactor is found to be equal to 17 for 
the two estimations. }
\label{f.3b}
\end{figure}

The rheometer was operated in controlled shear rate mode. Figure 
\ref{f.3a} shows the shear stress \( \sigma  \) as a function of the shear
rate \( \dot{\gamma } \) for three different samples. At lower shear
rates, the stress increases linearly with the shear rate yielding
a constant viscosity \( \eta =\sigma /\dot{\gamma } \). Above
a certain shear rate, the gels undergo
a slight shear thickening. This is illustrated by a linear fit taken
to the low shear rates data for the 0.25 \% w/w borax sample in Figure 
\ref{f.3a}. The 
viscosity increase is of the order of several percents.

Upon further increasing the shear rate, a stress plateau {\it i.e.} a constant
stress \( \sigma _{p} \) over a range of shear rates is
observed. After the plateau, the stress increases again with the shear
rate but the viscosity remains shear-thinning. For even higher
shear rates, secondary flow instabilities due to the normal stresses,
{\it e. g.} the Weissenberg effect in the Taylor Couette cell, become significant
and the liquid expels itself from the measurement cell.

\section{Discussion}

The observation of a stress plateau is surprising: it indicates that 
for a given stress \( \sigma_p  \), different
shear rates, and consequently different viscosities exist in one and
the same system. As this is unlikely to happen for our material, the
most natural interpretation would be that, instead of two viscosities,
there are two possible boundary conditions. This effect is known as
\emph{stick-slip} behaviour and has frequently been associated with
the occurrence of stress plateaus in the engineering literature. In
agreement with the occurrence of stick-slip, we find that the measured
 viscosity oscillates with a large amplitude in the vicinity
of the plateau stress. This suggests that the oscillations observed for 
the fingers are also due to a stick-slip motion. 
Consistently, we observe that the critical tip velocity $U_c$ for the
onset of the instability in the ST experiments corresponds to a typical shear 
rate ($\sim U_c/b$) of the order $1s^{-1}$, which decreases with increasing
borax concentration. This shear rate value is thus compatible 
with the critical shear rate for the occurrence of the stress plateau in
the rheology measurements for the different gels (Figure \ref{f.3a}). 
In order to further test this hypothesis, we compare
the observed wavelength of the oscillating fingers with the results
of a linear stability analysis for visco-elastic fingering.

Recent theory and experiments\cite{lindner1} on the instability in
visco-elastic (yield stress) fluids has shown that a linear stability
analysis of the flat interface can be performed; the wavelength of
maximum growth follows as: \( \lambda =\sqrt{\gamma b/\sigma_w} \),
with \( \sigma _{w} \) the wall shear stress, \( \gamma  \) the
surface tension and \( b \) the plate spacing of the Hele-Shaw cell.
For yield stress fluids, it was demonstrated that the wavelength at
early times after the instability verifies this prediction with the
wall shear stress equal to the yield stress plateau. In a similar way,
 for our visco-elastic fluid, the wall shear stress in the stick-slip regime
should be given by the plateau value $\sigma_p$. Under this assumption, 
one would then expect the oscillation wavelength $\lambda$ to write:
\begin{equation}
\label{equation1}
\lambda =\sqrt{\gamma b/\sigma _{p}}
\label{equation}
\end{equation}

This has a number of interesting consequences. First, it explains
why the characteristic wavelength turns out to be independent of the velocity. The other consequences
are that the finger width should scale as the square root of the
plate spacing and be independent of the channel width. We actually
observed that doubling the depth $b$ of the channel increases
 \( \lambda \)  by a factor of about 1.15 (instead of the expected $\sim$ 1.4 
). However, a similar increase was found when doubling the width
$W$ of the channel. In this geometry with a low $W/b$ ratio, the lateral 
confinement also plays a significant role in selecting the oscillation 
wavelength.

Figure \ref{f.3b} compares this prediction of $\lambda$
(expression (\ref{equation1})) to the experimentally
observed oscillation wavelength. We find a similar dependence of both data
on the plateau stress $\sigma_p$ obtained from rheology measurements; 
if we use the surface tension of water (72
mN.m\( ^{-1} \)) as the surface tension of our aqueous polymeric fluid, the
proportionality constant is found to be equal to 17.

That the physical origin of the instability is elastic follows directly
if we determine the elastic modulus from rheology measurements in oscillation mode. The
measurements indicate that the system has more than a single relaxation
time\cite{robb}. We therefore determine 
the elastic modulus $G_0$
as the point where the storage modulus $G'$ equals the loss modulus
$G''$; this is exact for a  Maxwell fluid, and should be a
good approximation for our fluid. From these experiments it follows that the elastic modulus $G_0$ and
the plateau stress are indistinguishable within the experimental error. Therefore, we obtain an equally good description of the data in Fig. \ref{f.3b} when plotting \( \lambda =\sqrt{\gamma b/G_{0}} \), with the same
proportionality constant.

Beyond the oscillating regime, the onset of the tadpole regime 
can be understood as a transition to a ``soft fracture'' dynamics.
This regime appears to be qualitatively very different from the viscous
one. In standard ST experiment, the injected air displaces a large
volume of liquid, inducing dissipation in the whole region ahead of
the tip. By contrast, in this fast regime,
the total flux of liquid becomes extremely small and the 
viscous dissipation is confined to a very small volume around the tip (as 
probed by the absence of motion of air bubbles or dust that
can be trapped between the tip and the outlet).  
If we evaluate the typical time-scale $\tau_p$ for the passage of the head 
(its length divided by the tip velocity)
we find $\tau_p<<\tau_r$  where $\tau_r \sim 1s$ is the relaxation 
time of the gel deduced from rheology measurements.
 Therefore, most of the gel
experiences an elastic deformation which is mainly recovered after the tip
has moved by. It is therefore qualitatively similar
to a fracture type dynamics. The unusual rounded shape of the  
tip is due to the low value of the elastic modulus $G_0$ for this material
(of order $100-500 Pa$). By comparing
elastic to surface forces, one can access a typical lengthscale 
$h_0=\gamma/G_0$ for the curvature of the tip  
of order $1 mm$. This value is consistent with
 values measured in the experiments.

\section{Conclusion}

In conclusion, we have observed an oscillating growth dynamics occurring upon
injection of an air bubble into a visco-elastic gel. This result is qualitatively
different from what had been observed so far in ST experiments with non-Newtonian 
liquids. By contrast with other systems where the elastic characteristic 
 of the invaded phase was found to increase tip splitting\cite{vlad}, 
here the finger experiences an 
instability along its growth direction. As a result, the thickness of the 
deposited film exhibits strong but spatially regular modulations.

Apart from geometrical parameters, we found the oscillation wavelength 
to be entirely controlled by the visco-elastic property of the gel, and to be
 independent of the injection rate. We think that this observation may be helpful, for instance, 
to understand the shear-induced emulsification of a dispersion of 
inviscid droplets in a visco-elastic matrix: it has been found that
such a combination could lead to very monodisperse emulsions\cite{bibette}. The final droplets 
size could result from a similar process occurring during the rapid elongation 
of the drops in the visco-elastic phase.  

\acknowledgments
We would like to thank J. Meunier, P. Sens and J.F. Joanny for fruitful discussions. LPS de l'ENS is UMR 8550 of the CNRS, associated with the Universities Paris 6 and 7.




\acknowledgments


\begin{thebibliography}{0}

\bibitem{st}
   \Name{P.G. Saffman \and G.I. Taylor}
   \REVIEW{Proc. R. Soc. London, Ser. A}{245}{1958}{312}

\bibitem{maher}
   \Name{K.V. McCloud \and J.V. Maher}
   \REVIEW{Physics Reports}{260}{1995}{260}{139}

\bibitem{coussot}
   \Name{P. Coussot}
   \REVIEW{J. Fluid Mech.}{380}{1999}{363}

\bibitem{lindner1}
   \Name{A. Lindner, P. Coussot \and D. Bonn}
   \REVIEW{Phys. Rev. Lett.}{85}{2000}{314}

\bibitem{lindner2}
   \Name{A. Lindner, D. Bonn \and J. Meunier}
   \REVIEW{Physics of Fluids}{12}{2000}{256}

\bibitem{shelley}
   \Name{L. Kondic, M. Shelley \and P. Pallfy-Muhoray}
   \REVIEW{Phys. Rev. Lett.}{80}{1998}{1433}

\bibitem{zhao}
   \Name{H. Zhao \and J. V. Maher}
   \REVIEW{Phys. Rev. E}{47}{1993}{4278}

\bibitem{vlad}
   \Name{D. H. Vlad \and J. V. Maher}
   \REVIEW{Phys. Rev. E}{61}{2000}{5439}

\bibitem{garlick}
   \Name{M. Garlick}
   http://www.delta.edu/slime/slime.html

\bibitem{chen}
   \Name{Yuan Chen Chung \and Tzyy-Lung Yu}
   \REVIEW{Polymer}{38}{1997}{2019}

\bibitem{robb}
   \Name{I.D. Robb \and J.B.A.F. Smeulders}
   \REVIEW{Polymer}{38}{1997}{2165}

\bibitem{bibette}
   \Name{T. G. Mason \and J. Bibette}
   \REVIEW{Phys. Rev. Lett.}{77}{1996}{3481}


\end{thebibliography}
\end{document}